\documentclass[manuscript,screen]{acmart}

\AtBeginDocument{%
  }

\setcopyright{acmlicensed}
\copyrightyear{2025}
\acmYear{2025}
\acmDOI{XXXXXXX.XXXXXXX}

\acmConference[Conference acronym 'XX]{X}{April 26, 2025}{XXXX}
\acmISBN{XXX}

\usepackage{bm}
\usepackage{multirow}
\usepackage{subfig}
\usepackage{algorithmic}
\usepackage{amsfonts}
\usepackage{booktabs}
\usepackage{graphicx}
\usepackage{comment}
\usepackage{hyperref}
\usepackage{fancyvrb}
\usepackage{cancel}
\usepackage{amsmath}

\newcommand{\ExactPhase}{Exact Match}
\newcommand{\SimilarPhase}{Similar Match}
\newcommand{\InversePhase}{Inverse Match}

\title{The Influence of Text Variation on User Engagement in Cross-Platform Content Sharing}

\author{Yibo Hu}
\affiliation{%
  \institution{Georgia Institute of Technology}
  \country{USA}}
\email{yibo.hu@gatech.edu}

\author{Yiqiao Jin}
\affiliation{%
  \institution{Georgia Institute of Technology}
  \country{USA}}
\email{yjin328@gatech.edu}

\author{Meng Ye}
\affiliation{%
  \institution{SRI International}
  \country{USA}}
\email{meng.ye@sri.com}

\author{Ajay Divakaran}
\affiliation{%
  \institution{SRI International}
  \country{USA}}
\email{ajay.divakaran@sri.com}

\author{Srijan Kumar}
\affiliation{%
  \institution{Georgia Institute of Technology}
  \country{USA}}
\email{srijan@gatech.edu}

\begin{document}


\begin{abstract}
In today's cross-platform social media landscape, understanding factors that drive engagement for multimodal content, especially text paired with visuals, remains complex. 
This study investigates how rewriting Reddit post titles adapted from YouTube video titles affects user engagement.
First, we build and analyze a large dataset of Reddit posts sharing YouTube videos, revealing that 21\% of post titles are minimally modified. Statistical analysis demonstrates that title rewrites measurably improve engagement. 
Second, we design a controlled, multi-phase experiment to rigorously isolate the effects of textual variations by neutralizing confounding factors like video popularity, timing, and community norms. Comprehensive statistical tests reveal that effective title rewrites tend to feature emotional resonance, lexical richness, and alignment with community-specific norms.
Lastly, pairwise ranking prediction experiments using a fine-tuned BERT classifier achieves 74\% accuracy, significantly outperforming near-random baselines, including GPT-4o. These results validate that our controlled dataset effectively  minimizes confounding effects, allowing advanced models to both learn and demonstrate the impact of textual features on engagement.
By bridging quantitative rigor with qualitative insights, this study uncovers engagement dynamics and offers a robust framework for future cross-platform, multimodal content strategies.

\end{abstract}

\ccsdesc[500]{Human-centered computing~Social networks}
\ccsdesc[500]{Computing methodologies~Natural language processing}

\keywords{cross-platform content analysis, text mining, user engagement modeling, cross-modal social media}

\maketitle

\section{Introduction}

Social media platforms today host diverse content types—text, images, videos, and audio—shared widely across interconnected networks. This cross-platform sharing of multimodal content profoundly shapes user engagement, making it critical to understand the factors driving success  \cite{Diaz14_twitter_cr,Abdollahi14_rank_twitter,aldous2019view,stoddard2015popularity}. 
Platforms like Reddit and YouTube exemplify this interconnected dynamic, where users adapt and redistribute content under varying community norms \cite{AnatomyOfReddit,RMNReddit,ConvoModeling,CommentPop2,CommentPop1,CommentPop3,vallet2015characterizing,abisheva2014watches,soysa2013predicting,buntain2021youtube,gkikas2022text}. However, identifying the factors driving post success in this complex environment remains a significant challenge for researchers and practitioners \cite{FactorsInPop2,ReinforcePopPredict}.

While prior research has explored cross-platform \cite{jin2023predicting} and multimodal  content-sharing dynamics \cite{nguyen2024supporters, verma2024community}, limited attention has been given to how textual variations influence engagement in such contexts. This challenge is compounded by the dominant influence of video popularity, which often overshadows textual factors. Addressing this gap requires disentangling textual effects from other variables in cross-platform sharing scenarios.

\begin{figure}
    \centering
    \includegraphics[width=0.5\linewidth]{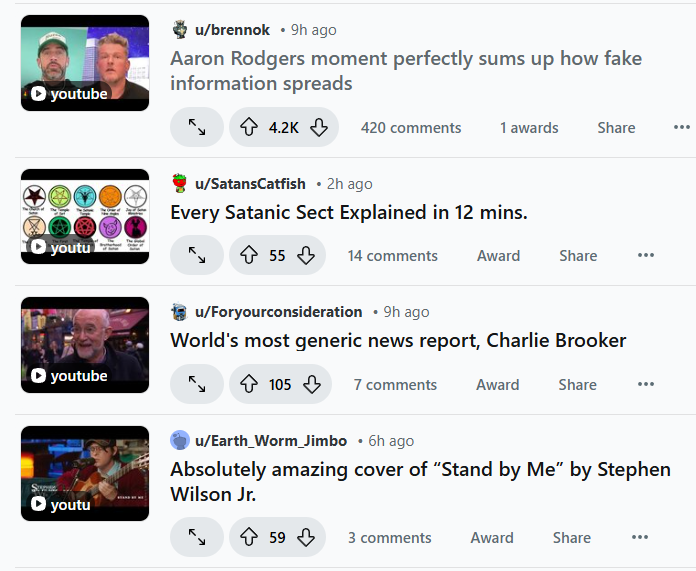}
    \caption{Example posts from \texttt{r/videos}, Reddit’s largest video-sharing subreddit. 
The ranking and visibility of posts are influenced by user votes, titles, video content, timing, and interaction dynamics.}
    \label{fig:subreddit_snapshot}
    \Description{This figure shows a list of example posts from the subreddit r/videos. Each post displays its title, number of comments, upvotes, and sharing platform (YouTube). It highlights how engagement is distributed across various titles, reflecting differences in scores based on title characteristics, timing, and user interactions.}
\end{figure}

This study focuses on textual variations in Reddit post titles adapted from YouTube videos and their impact on user engagement.  
\textbf{Titles} represent a low-effort yet high-impact strategy to align  content with  audience preferences, potentially driving significant changes in engagement.
Despite studies on titles’ impact in news and social media \cite{piotrkowicz2017impact,kuiken2017effective_headline, park2021present, guo2022verba_Title_Changes,weissburg2022judging}, their role in cross-platform video sharing remains underexplored.  By controlling for factors like video popularity, this work isolates the specific impact of text rewrites on engagement.

We focus on \textbf{Reddit} due to its unique community-driven structure, where topic-focused communities (\textit{subreddits}) foster distinct content norms. Unlike  individual-centric platforms like  Twitter or Instagram, Reddit reduces author-related influence \cite{weissburg2022judging}, allowing us to isolate the impact of title variations on engagement across diverse audiences. Figure \ref{fig:subreddit_snapshot} highlights these dynamics through example posts from \texttt{r/videos}, Reddit’s largest video-sharing subreddit, where post titles, video content, and user interactions interplay to shape engagement.

This study addresses three key questions: 
1) How do textual variations in online posts influence engagement in cross-platform video sharing?
2) How can confounding factors like timing, video popularity, and subreddit norms be disentangled to isolate the impact of textual variations? 
3) What engagement patterns emerge within specific Reddit communities regarding title effectiveness?

To answer these questions, we build a large-scale dataset of Reddit posts sharing YouTube videos to enable a comprehensive cross-platform, multimodal analysis. 
Preliminary analysis reveals that 21\% of post titles are minimally modified,  highlighting a clear opportunity for title rewrites to boost engagement. Statistical analysis further demonstrates that such rewriting has a measurable and positive impact on user engagement.

To rigorously measure the impact of title rewrites, we design a multi-phase, controlled experiment. In the \textbf{\ExactPhase{} Phase}, we examine post pairs sharing the same video to study the timing effect. The \textbf{\SimilarPhase{} Phase} expands the analysis to post pairs sharing similar videos with comparable view counts, isolating  title effectiveness from video popularity. 
Finally, the \textbf{\InversePhase{} Phase} focuses on cases where a more popular post shares a less popular video, further controlling for video influence to highlight text-based impact. 

Through this incremental analysis, we leverage advanced deep learning models to supplement traditional lexical-based approaches, capturing subtle differences that reveal patterns in title effectiveness. For instance, popular posts in controlled settings often feature longer titles, sentiment shifts, or keywords that resonate with community norms.
These findings are rigorously validated using both parametric and non-parametric statistical tests.

To further validate our findings, we conduct pairwise ranking prediction experiments under these controlled settings. Heuristic baselines, such as random guessing and time-based ranking, performed near-randomly, affirming that the controlled dataset effectively isolates text-based effects. GPT-4o, utilized as an automated evaluator, shows only a slight improvement over random guessing, reflecting the same limitations faced by human judgment in predicting post popularity based on limited textual cues. In contrast, a fine-tuned BERT classifier achieves  74\% accuracy, significantly outperforming baselines and underscoring the importance of textual features for engagement. These results show that our controlled setting effectively overcomes the inherent complexity of social network data. By isolating textual effects, adapted advanced models can learn and detect subtle patterns that drive engagement, as corroborated  by our statistical findings.
Although predictive accuracy is not our sole goal, these findings highlight the tangible influence of title rewrites on post popularity, offering insights into cross-platform content-sharing dynamics.
In summary, our contributions are as follows:
\begin{itemize} 
    \item \textbf{Novel Problem.} Examining how textual variations in post titles influence engagement in cross-platform, multimodal content sharing.
    \item \textbf{Novel Dataset.} A large-scale dataset of Reddit posts sharing YouTube videos, with detailed analysis of title modifications and engagement metrics. 
    \item \textbf{Comprehensive Evaluation.} A multi-phase controlled experiment combined with prediction tasks to disentangle text-based effects from confounding factors, providing actionable insights for cross-platform content strategies.
\end{itemize}

Together, these contributions advance cross-platform social media research by offering insights into community-specific engagement strategies and establishing a replicable framework for optimizing multimodal content.

\section{Related Work}\label{sec:related work}

\subsection{Observational Studies on Engagement} 

Research on social media engagement prediction has primarily focused on content-based and network-based factors. 
While \textbf{network-based approaches} analyze user relationships and social dynamics (e.g., follower networks or collaborative filtering) \cite{Diaz14_twitter_cr, Abdollahi14_rank_twitter, he2017neural,Volkovs20_twitter_predict_LM, Felicioni20_twitter_predict, Purohit21_twitter_features, Daniluk21_twitter_predict_neural,arazzi2023predicting,jin2023code,jin2023predicting,jin2024empowering}, they do not directly examine how textual features influence engagement, which is the focus of our study.

In contrast,  \textbf{content-based methods} leverage features such as user profiles, engagement metrics (e.g., likes, retweets), and, most importantly, textual attributes \cite{Chen12_twitter_in_occupy_Wall_Street, Magalhaes14,Orellana16,  Silva20_predict_twitter_covid19_engagement, Toraman22_social_twitter, AlAnJa22,he2025survey}.  
Numerous studies highlight the pivotal role of text in shaping user engagement. 
For example, \citet{aldous2019view} demonstrated that content type and context significantly influence user interactions across five platforms. 
Similarly, \citet{Toraman22_social_twitter} found that tweet engagement is primarily driven by text semantics rather than author popularity \cite{silva2018analyzing, Volkovs20_twitter_predict_LM}.
Key features such as readability \cite{gkikas2022text} and sentiment \cite{saquete2022some} further influence user responses.
These findings establish textual content as a primary determinant of engagement, reinforcing the relevance of our focus on title rewrites in cross-platform scenarios.

\subsection{Controlled Studies on Engagement}

Predicting social media engagement is inherently challenging due to the multivariate nature of online content and the feedback loops in user reactions \cite{FactorsInPop2,ReinforcePopPredict,FactorsInPop3,PairwisePopPredict1,FactorsInPop1}. 
Controlled experiments help isolate the impact of individual factors \cite{salganik2006experimental,stoddard2015popularity}. 
For instance, \citet{Lakkaraju2013_TitlePopPredict} analyzed images posted on various subreddits with different titles and found that while effective titles boost engagement, timing effects often overshadow their influence. Similarly, \citet{Tan14-Topic-and-author-controlled-Twitter} compared tweets with matched topics and timing, revealing that text alone moderately predicts engagement, while human judgments offer only slight improvements over random guessing.
\citet{CatsAndCaptions} extended this approach by analyzing the impact of captions and images on Reddit engagement using a pairwise ranking task that controls for timing and community effects. Building on this, \citet{weissburg2022judging} demonstrated that titles alone can predict Reddit engagement, uncovering community-specific patterns that affect popularity.

Related work on news headline editing has also explored title modifications to improve engagement on platforms like Twitter and Facebook \cite{piotrkowicz2017impact,piotrkowicz2017headlines,kuiken2017effective_headline, park2021present, guo2022verba_Title_Changes}. These studies emphasize the potential of textual variations in driving user interactions but are limited to single-platform contexts.

While these studies provide insights into single-platform engagement, our study addresses the added complexity of cross-platform interactions and multimodal factors like video content.

\subsection{Causal Inference for Content Engagement Analysis}
Causal inference methods offer another approach to isolate confounding factors in engagement prediction. Techniques like LDA, auto-encoders, and propensity score matching have been used to assess the impact of text properties \cite{pryzant2017predicting, pryzant2020causal,fytas2021makes,li2023boosts}. For example, \citet{park2021present} used propensity score matching to analyze how clickbait editing styles impact engagement.
However, adapting causal inference to a multimodal, cross-platform setting introduces challenges, such as dependencies between video and text. Our study focuses on isolating textual effects using a controlled setup, leaving the integration of multimodal causal inference for future exploration.

\subsection{Cross-Platform and Comparative Social Media Studies} 

Cross-platform studies often focus on predicting video popularity through network-driven predictors like tweet rates, shares, and follower counts \cite{vallet2015characterizing,soysa2013predicting,abisheva2014watches, yu2014twitter},  rather than directly examining engagement. 
Separately, \citet{arazzi2023importance} analyzed linguistic evolution across Twitter and Reddit but treated each platform in isolation.
In contrast, our study bridges this gap by analyzing Reddit title-based engagement for YouTube video shares. By isolating textual features through controlled experiments, we emphasize the role of title variations in driving engagement, moving beyond network-centric approaches.

\section{Dataset Construction and Analysis}\label{sec:dataset}

\subsection{Datasets Overview and Integration} 

Our dataset comprises the following components:

\begin{itemize}
    \item \textbf{YouTube Video Metadata:} Metadata for 12 million YouTube videos, including titles, authors, descriptions, channels, topics, and tags.
    \item \textbf{Reddit-YouTube Propagation Posts:} A five-years collection of Reddit posts from 2018 to 2022 that share YouTube videos, totaling 59.6 million shares from 26 million unique videos.  
    These posts primarily fall into two categories:
    \begin{itemize}
        \item \textbf{Video Posts:} The majority of posts embed YouTube videos, with the title as the only textual element.
        \item \textbf{Text Posts:} A smaller subset with linked videos in the post body.
    \end{itemize}
\end{itemize}

We focus primarily on video posts for their data volume and suitability for analyzing title rewrites without interference from additional textual content.  
Text posts, though limited, provide supplementary data for broader context.

To integrate these datasets,  we standardized YouTube URLs\footnote{\texttt{www.youtube.com/watch?v=[video\_id]}} to extract unique Video IDs, retaining only direct video links while excluding channels, playlists, or user profiles.
This produced a refined dataset of 24.1 million posts (19.8M video posts,  4.3M text posts) across 305K subreddits and 12.6 million unique videos.


By combining metadata with post information, our dataset facilitates a nuanced study of title-driven engagement within subreddit-specific contexts. We focus on the role of titles while considering three critical confounding variables: subreddit factors, post timing, and video characteristics. In the following subsections, we examine these variables to identify their influence on engagement, setting the stage for the controlled analysis in Section \ref{sec:Controlled Analysis}.

\subsection{Subreddit Factors}

\paragraph{Subreddit Metrics and Engagement Trends}
Engagement within subreddits is highly imbalanced, driven by a few dominant communities.  Subreddit user counts (i.e., community size) follow a power-law distribution ($\alpha$=1.84, $\sigma$=0.013), while post mean scores\footnote{In this paper, ``posts'' and related metrics like counts and mean scores apply exclusively to Reddit posts sharing YouTube videos, rather than arbitrary post types.} show a similar skew ($\alpha$=1.95, $\sigma$=0.002). 
Table \ref{tab:example_subreddits} illustrates these dynamics with examples from the top 10 subreddits by post count. For example, \texttt{r/GetMoreViewsYT} shows low engagement, while \texttt{r/videos} has a skewed distribution with high mean but low median scores.  In contrast, \texttt{r/kpop} and \texttt{r/SquaredCircle} show consistent engagement with higher median scores.

\begin{table}[tbp]
\centering
\caption{Top 10 subreddits ranked by the number of video-sharing posts, showing metrics such as the number of users, total posts, and mean/median post scores. The table highlights significant variations in engagement across subreddits. }
\label{tab:example_subreddits}
\begin{tabular}{ccccc}
\toprule
\multirow{2}{*}{\textbf{Subreddit}} & \multirow{2}{*}{\textbf{\# Users}} & \multirow{2}{*}{\textbf{\# Posts}} & \multicolumn{2}{c}{\textbf{Post Scores}}    \\
    &    &    & \multicolumn{1}{c}{\textbf{Mean}} & \multicolumn{1}{c}{\textbf{Median}}  \\
\midrule
r/videos    & 26.9M    & 350K    & 305.9    & 1.0    \\
r/Music    & 33.0M    & 258K    & 63.9    & 2.0    \\
r/GetMoreViewsYT    & 112K    & 189K    & 1.4    & 1.0    \\
r/YouTube\_startups    & 109K    & 178K    & 1.3    & 1.0    \\
r/SmallYoutubers    & 88K    & 158K    & 1.2    & 1.0    \\
r/kpop    & 2.7M    & 114K    & 108.5    & 52.0    \\
r/asmr    & 282K    & 81K    & 9.1    & 3.0    \\
r/listentothis    & 18.0M    & 78K    & 37.8    & 3.0    \\
r/leagueoflegends    & 7.1M    & 53K    & 217.2    & 2.0    \\
r/SquaredCircle    & 872K    & 50K    & 82.0    & 17.0    \\
\bottomrule
\end{tabular}
\end{table}

Further analysis reveals that community size strongly affects engagement in smaller subreddits but has a diminishing impact in those with over 100K users. Similarly, post counts do not consistently boost engagement in larger subreddits. Table \ref{tab:correlation_analysis} quantifies these trends: a moderate Pearson correlation (0.311) between user count and mean post scores suggests that larger subreddits may foster engagement, but a low Spearman correlation (0.105) reveals inconsistencies due to outliers and non-linear dynamics. Additionally, weak correlations between post counts and mean scores imply that higher posting volumes alone do not guarantee greater engagement.

To reduce noise from smaller, less active communities and focus on robust engagement patterns,  we selected the top 5,000 subreddits, representing the largest 1-2\% as of February 13, 2024. Each included subreddit features at least 1,000 valid video-sharing posts, resulting in a refined dataset of 3,043 subreddits. 

\begin{figure}[bp]
    \begin{minipage}{0.3\linewidth}
        \centering
    \includegraphics[width=\linewidth]{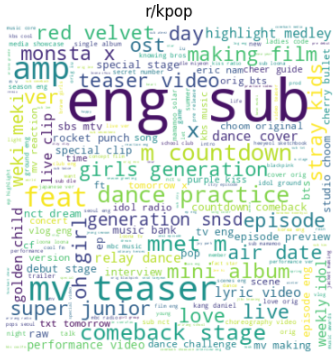}
    \end{minipage}%
    \hspace{0.02\linewidth}
    \begin{minipage}{0.3\linewidth}
    \centering
    \includegraphics[width=\linewidth]{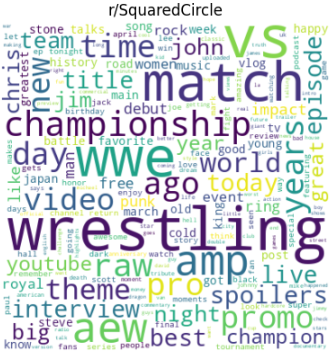}
    \end{minipage} %
    \hspace{0.02\linewidth}
    \begin{minipage}{0.3\linewidth}
    \centering
    \includegraphics[width=\linewidth]{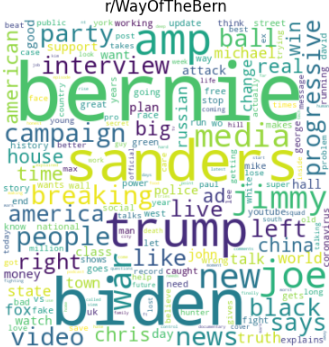}
    \end{minipage} 
    \caption{Word clouds illustrate the distinct thematic focuses of video posts in three large subreddits:  \texttt{r/kpop}, \texttt{r/SquaredCircle}, and \texttt{r/WayOfTheBern}, highlighting their unique community norms.}
    \label{fig:subreddit_word_cloud}
    \Description{This figure contains three word clouds illustrating the thematic focuses of video posts in three large subreddits: r/kpop, r/SquaredCircle, and r/WayOfTheBern. The word clouds highlight commonly used terms that align with the interests of each subreddit, such as "teaser" in r/kpop and "wrestling" in r/SquaredCircle.}
\end{figure}

\paragraph{Community-Specific Norms.}
Beyond quantitative metrics, community norms shape engagement through thematic preferences, stylistic conventions, and audience expectations.  Figure \ref{fig:subreddit_word_cloud} illustrates this with word clouds from three large subreddits: \texttt{r/kpop} focuses on music and performances, \texttt{r/SquaredCircle} centers on wrestling, and \texttt{r/WayOfTheBern} highlights political discourse.   While filtering out smaller subreddits improves quantitative analysis, community norms remain significant. Isolating these factors ensures that the impact of title-level features can be accurately assessed without interference from broader subreddit-specific dynamics.

\begin{table}[tbp]
\centering
\caption{Correlation between subreddit-level metrics, showing the relationships among user numbers, post counts, and mean post scores for video-sharing subreddits.}
\label{tab:correlation_analysis}
\begin{tabular}{@{}l@{\hspace{6pt}}c@{\hspace{6pt}}lcc@{}}
\toprule
 &  &   &  \textbf{Pearson}  & \textbf{Spearman} \\ 
 \midrule
\# Users & vs. & \# Posts  & 0.247 & 0.205 \\
\# Users & vs. & Mean Post Score  & 0.311 & 0.105 \\
\# Posts & vs. &  Mean Post Score & 0.018 & 0.139 \\ 
\bottomrule
\end{tabular}
\end{table}

\begin{figure}
    \centering
    \includegraphics[width=1\linewidth]{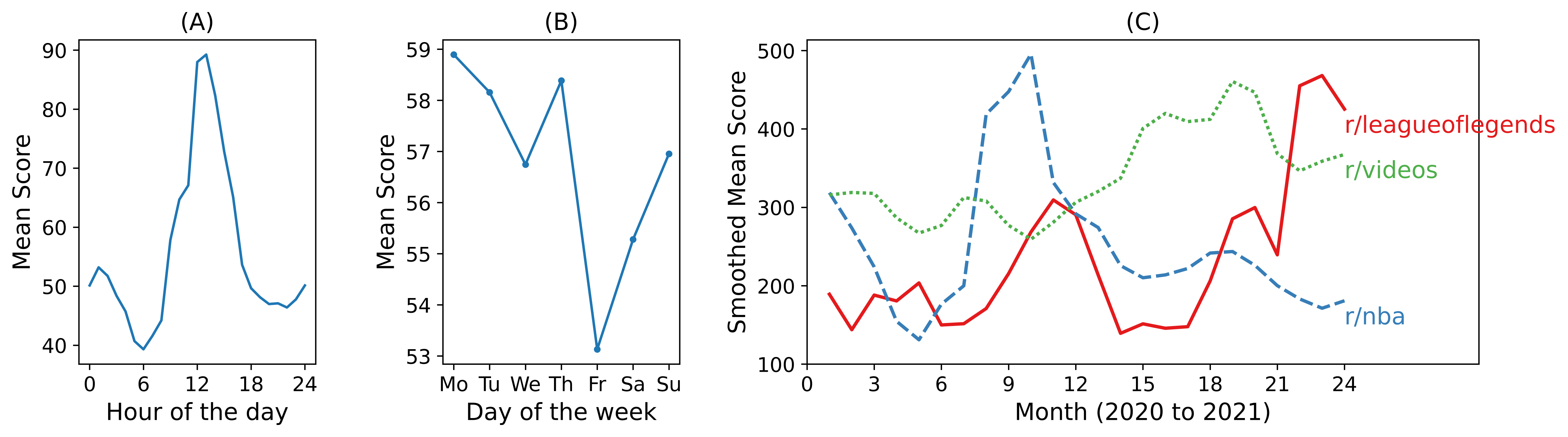}
    \caption{Temporal variation in mean post scores by (A) hour of the day, (B) day of the week, and (C) month (smoothed over three-month intervals for three subreddits during 2020-2021)}
    \label{fig:time_factors}
    \Description{This figure presents three line charts. (A) shows the variation in mean post scores by the hour of the day. (B) displays mean post scores by the day of the week, indicating engagement trends over time. (C) shows mean post scores by month }
\end{figure}

\subsection{Post Timing}
Post engagement fluctuates across daily and weekly cycles, with distinct peaks and troughs (Figure \ref{fig:time_factors}).  Over longer timescales, engagement trends reflect broader influences such as trending topics and major events \cite{CatsAndCaptions}. 
For instance, (C) illustrates smoothed monthly mean scores for three subreddits during 2020-2021, which highlight significant fluctuations in popularity. 
On shorter timescales, posts published earlier may achieve higher engagement due to longer exposure. 
These patterns underscore the importance of controlling for timing in engagement analysis.


\subsection{Video Characteristics}
\begin{figure}[htbp]
    \centering
    \includegraphics[width=1\linewidth]{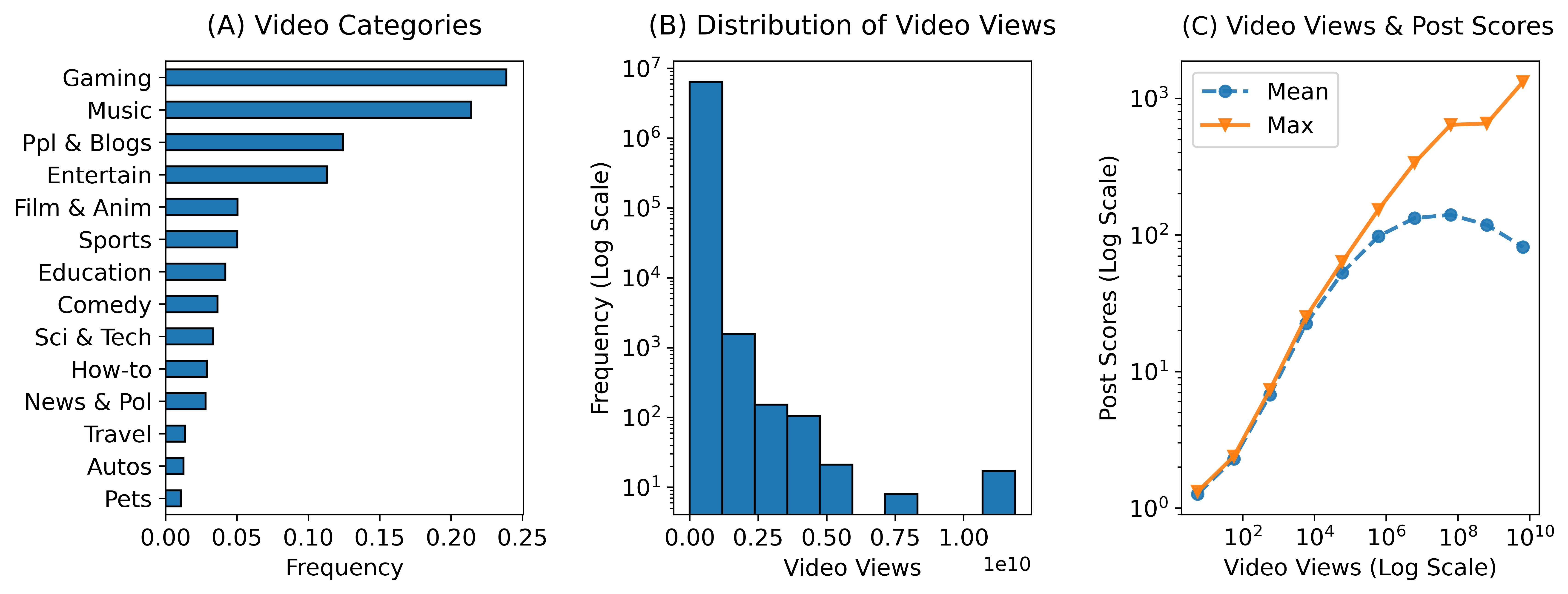}
    \caption{Video Statistics: (A) Frequency distribution of YouTube video categories; (B) Log-scaled distribution of video views; (C) Positive correlation between video views (log scale) and post scores, with mean scores peaking at around $10^7$ views.}
    \label{fig:video_summary}
    \Description{This figure consists of three subplots: (A) Bar chart showing the frequency of YouTube video categories, with gaming and music being the most popular. (B) Histogram showing the highly skewed distribution of video views on a logarithmic scale. (C) Scatter plot illustrating the positive correlation between video views (log scale) and mean/max post scores.}
\end{figure}

We analyze two key aspects of video characteristics: topics and popularity, as shown in Figure \ref{fig:video_summary}.

\paragraph{Video Topics}
YouTube videos are classified with \textbf{tags} and \textbf{categories}. 
Each video has multiple detailed tags (e.g., ``Pop Music,'' ``Action Game'') and can be grouped into four broad groups: Music (29.0\%), Games (27.5\%), Society/Lifestyle (26.8\%), and Entailment (11.8\%). Ambiguous tags such as Video (26.8\%) and minor topics  (7.1\%) that overlap with other subjects were excluded for clarity. This grouping outlines the diverse interests that drive video sharing on Reddit. In contrast, Youtube categories provide a single, high-level classification for each video. Our dataset includes 14 categories (Figure \ref{fig:video_summary} (A)), which complement tag-based groupings.  For instance, the ``Sports'' category aligns with the broader Entailment group.  We primarily use categories for video topic matching as they balance tag granularity and simplicity.

\paragraph{Video Popularity}
Video views, a key indicator of popularity, follow a power-law distribution ($\alpha$=1.11, $\sigma$=4e-5) (Figure \ref{fig:video_summary} (B)). 
Logarithmic scaling reveals that higher views generally correspond to higher post scores. However, engagement plateaus for ultra-popular videos (over $10^7$ views), likely due to saturation effects from viral memes (Figure \ref{fig:video_summary} (C)).

\subsection{Post Titles}

After examining the primary confounding factors, we now focus on post titles themselves. 
Reddit video-sharing post titles show a slightly left-skewed distribution, with an average length of 10.3 words ($\sigma$ = 7.2).  While 95\% of titles are under 23 words, a minority extend up to 100 words, reflecting diverse style. To investigate the role of title modifications, we focus on two key questions:  (1) How frequently do users rewrite video titles? and (2) Do these modifications appear to impact engagement?

\begin{figure}[tbp]
    \centering
    \includegraphics[width=1\linewidth]{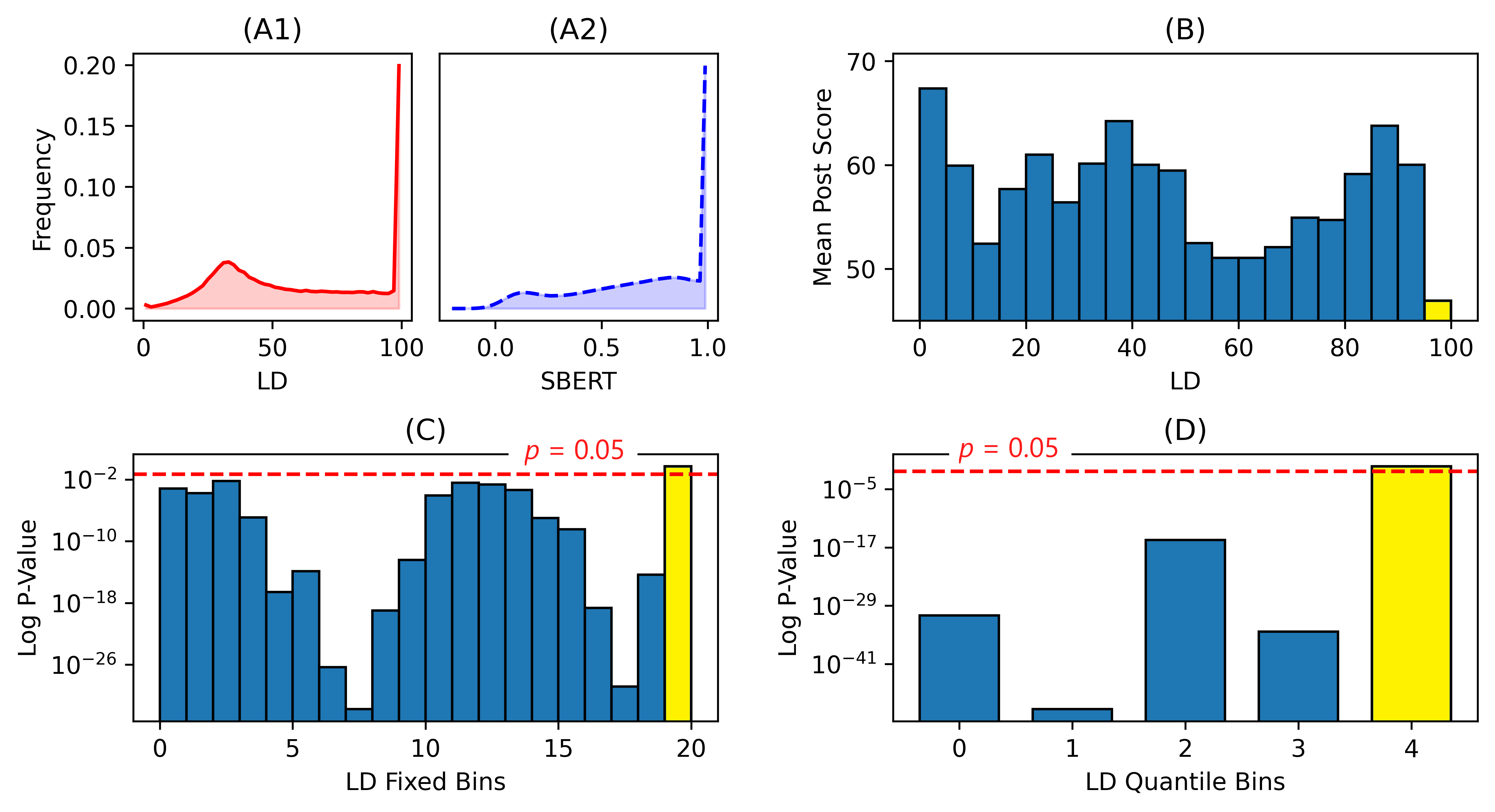}
        \caption{Title modifications and engagement: (A) Degree of rewriting in post titles compared to their corresponding video titles, measured using Levenshtein Distance (LD, A1) and SBERT cosine similarity (A2); (B) Mean post scores across LD groups, showing that rewritten titles lead to higher engagement than copied titles.  (C, D) Statistical significance (log-scale p-values) for fixed-width LD bins (C) and quantile-based LD bins (D), comparing each group to minimally modified titles (final bins, highlighted in yellow).}
        \label{fig:post_title}
        \Description{This figure shows: (A1) Histogram of Levenshtein Distance (LD) between post titles and corresponding video titles. (A2) Histogram of SBERT cosine similarity scores for the same. (B) A bar chart showing mean post scores for different LD groups. (C, D) Log-scale p-values for statistical significance of LD groups using fixed-width and quantile-based bins, indicating rewritten titles lead to higher engagement.}
\end{figure}

To quantify the extent of title rewriting, we compare post titles to their corresponding original video titles using two metrics, shown in Figure \ref{fig:post_title} (A).
\textbf{Levenshtein distance (LD)}, or edit distance,  measures the minimum single-character edits needed to change one title to another, normalized from 0 (completely different) to 100 (identical) to account for length differences\footnote{Normalized LD is implemented using the \texttt{fuzz.ratio} function from the python \texttt{fuzzywuzzy} library.}.  
Additionally, \textbf{SBERT cosine similarity} \cite{reimers-2019-sentence-bert} uses sentence embeddings to assess semantic changes. While LD measures surface-level changes, SBERT captures deeper meaning and is more right-skewed due to its robustness to minor edits.

Our analysis shows that \textbf{21\% of post titles remain nearly identical} to the original video title (LD $\geq$ 95). 
However, substantial modifications are strongly associated with higher engagement. As shown in Figure \ref{fig:post_title} (B),  lower LD values correlate with higher post mean scores, indicating that rewriting titles may enhance post popularity.

To validate these findings, we conduct statistical tests on LD bins.  
In Figure \ref{fig:post_title} (C), we divide the LD range into 20 \textbf{fixed-width bins}, each spanning 5 units (e.g., 0–5, 5–10). and compare the mean scores of each bin to a reference group (LD $\geq$ 95, the final bin in yellow)).  The tests confirm that bins with lower LD values significantly outperform the reference group, with $p<0.001$ in most cases, despite the latter has a larger sample size.

To address potential biases from uneven bin sizes, we conduct a supplementary analysis using \textbf{quantile-based bins} (Figure \ref{fig:post_title} (D)). The dataset is divided into five bins of roughly equal size (20\% each), with the final bin (LD $\geq$ 98) representing the reference group with minimally modified titles (highlighted in yellow).  This ensures comparable sample sizes across bins. The results reinforce the trend observed in the fixed-bin analysis, with all bins showing significant differences compared to the reference group  ($p\ll0.001$), further supporting the positive impact of substantial title modifications on engagement.

The findings also suggest that moderate title rewriting (e.g., LD around 40) is particularly effective for boosting engagement. While many users opt for minor changes, moderate adjustments appear to better balance the originality of the video title with the appeal of a rewritten headline.

However, the low Spearman correlation  (-0.05) between LD and post scores suggests individual variability and complex social dynamics. 
This calls for a more controlled framework to disentangle the effects of title modifications from other factors. 
The next section introduces a multi-phase experimental setup to systematically investigate how title modifications influence cross-platform engagement.

\section{Controlled Analysis of Title Effectiveness}\label{sec:Controlled Analysis}

\subsection{Multi-Phase Confound Control}

To isolate the effects of title rewriting on post popularity, we employ a multi-phase approach that systematically controls for confounding factors through paired posts. Unlike prior methods that paired identical posts \cite{Tan14-Topic-and-author-controlled-Twitter,Lakkaraju2013_TitlePopPredict} or broadly subreddit-level trends \cite{CatsAndCaptions,weissburg2022judging}, our framework addresses more complex confounders, particularly video popularity, in a progressive manner.  This balanced design captures nuanced title effects on engagement while preserving dataset size for robust analysis.  The analysis progresses through three experimental phases: \textbf{\ExactPhase{} (Exact)}, \textbf{\SimilarPhase{} (Similar)}, and \textbf{\InversePhase{} (Inverse)}, each controlling variables incrementally.

\paragraph{\ExactPhase{} Phase.} We analyze post pairs drawn from the same subreddit and share identical videos. By holding both the community and video content constant, this setup isolates the effects of posting time, with title variations treated as random. 

\begin{figure}[tbp]
  \centering
  \includegraphics[width=0.5\linewidth]{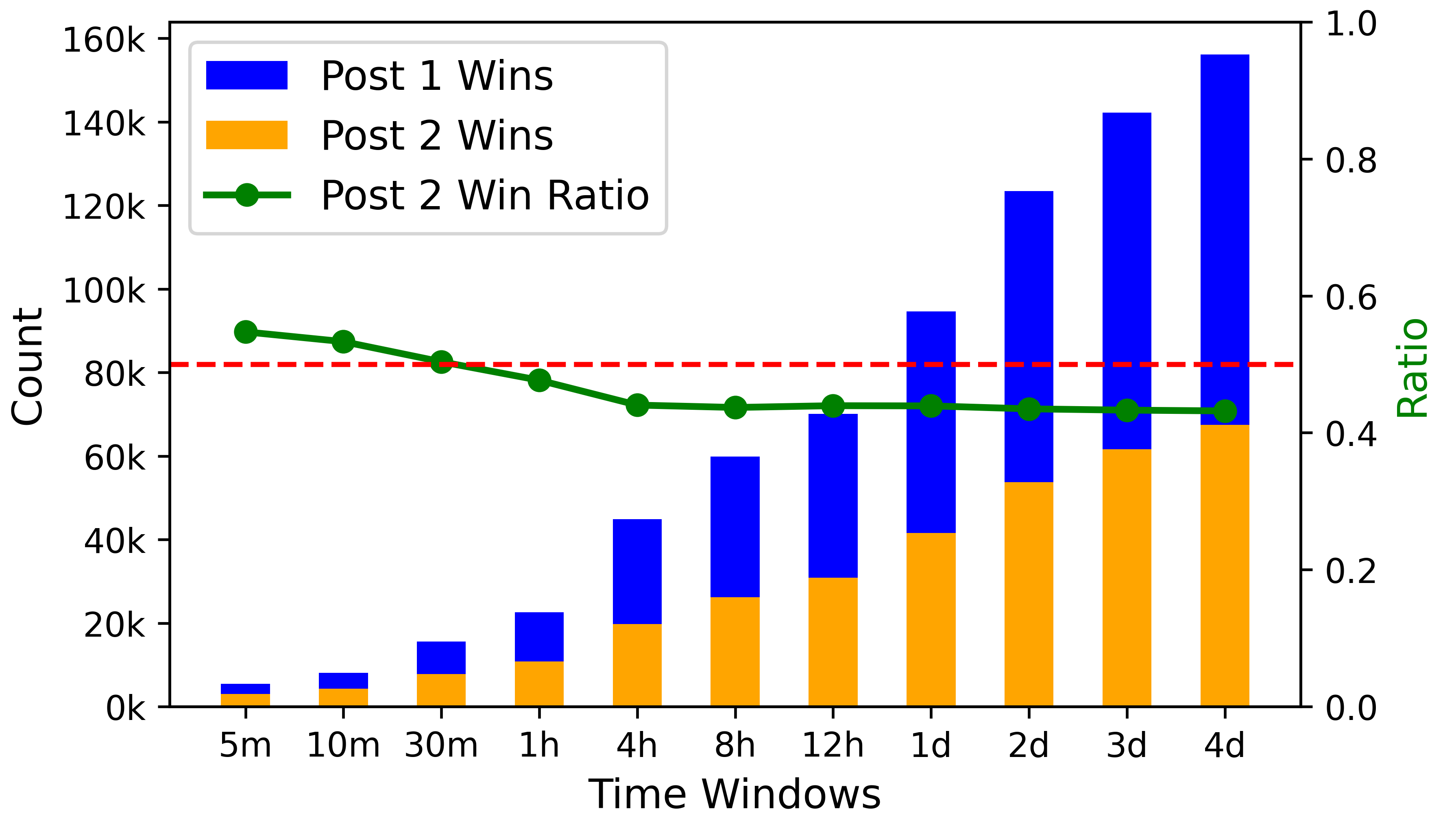}
  \caption{Time factor analysis in Exact Match Phase. Count and ratio of posts with higher scores across time windows, comparing Post 1 (earlier) and Post 2 (later) in a post pair. Time windows represent cumulative intervals (e.g., within 30 minutes or 1 hour).}
  \label{fig:post_time_window_5}
  \Description{This bar chart displays the count and ratio of superior posts (Post 1 vs. Post 2) across varying time intervals. The green line represents the ratio of posts where Post 2 performs better. The graph illustrates how timing influences post engagement dynamics.}
\end{figure}

Figure \ref{fig:post_time_window_5} shows engagement patterns across posting intervals.
Bars represent cases  where earlier posts (Post 1, blue) or later posts  (Post 2, yellow) perform better, while the green curve shows the percentage of cases where Post 2 outperforms Post 1. 
Posts with equal scores are excluded for better focus. 
A ratio of 0.5 suggests equal popularity between the two posts, whereas stabilization at 0.44 after four hours suggests that earlier posts generally gain more engagement due to longer exposure.  Short time windows (e.g., under 30 minutes) diminish this bias but reduce data availability.

Timing effects differ by post type: video posts, which refresh more frequently due to higher volume, exhibit shorter timing effects, while text posts retain visibility longer. 
These findings guide subsequent phases in selecting feasible time windows.

\begin{figure}[tbp]
    \centering
    \includegraphics[width=1\linewidth]{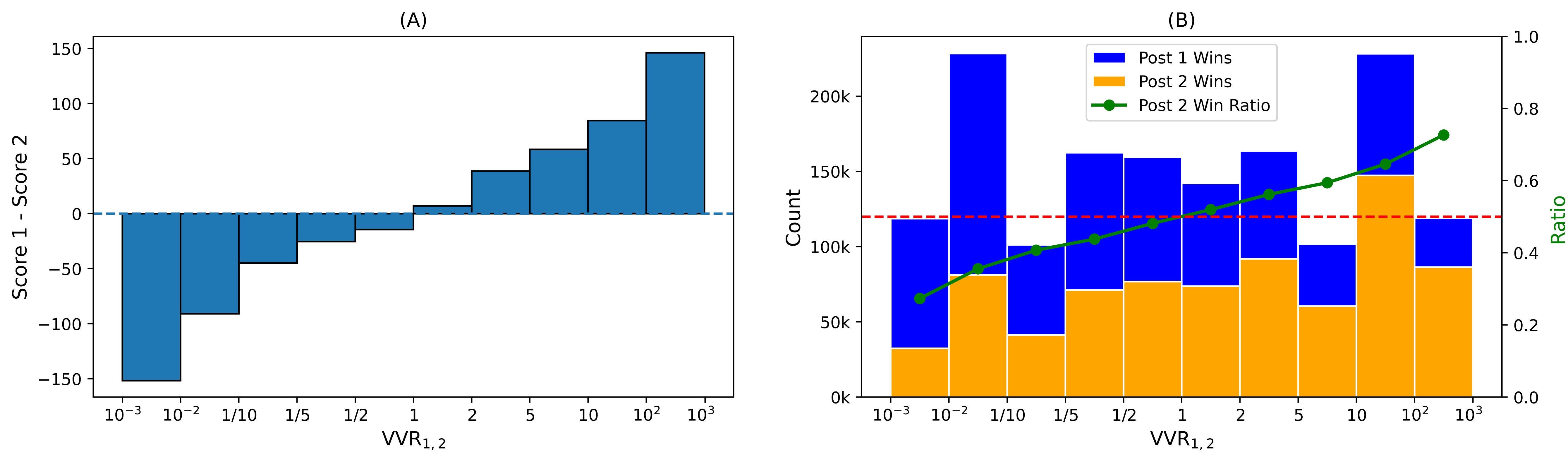}
     \caption{Video popularity analysis in Similar Match Phase. Post engagement across Video View Ratio (\(VVR_{1,2}\)) intervals (e.g., 1/2 to 1, 1 to 2), with Post 1 and Post 2 randomly ordered. (A) Mean score difference (Score 1 - Score 2) for each interval. (B) Count and ratio of posts by performance within each interval.  Both subfigures illustrate the strong relationship between VVR and post engagement.}
      \label{fig:score_views_interval}
      \Description{This figure comprises two subplots: (A) A bar chart displaying the mean score difference between Post 1 and Post 2 within each VVR interval. (B) Bar chart showing the count and ratio of superior posts across different video view ratios (VVR intervals). }
\end{figure}

\paragraph{\SimilarPhase{} Phase.} This phase expands the analysis to post pairs with similar videos, allowing us to study the influence of video popularity while treating titles as random. 
Posts are paired within the same subreddit and sampled within a 30-minute interval (based on the Exact Phase findings). Unlike the Exact Phase, these pairs share videos with the same categories and similar view counts.  
The Video Views Ratio (\textbf{VVR}), denoted as ($\text{VVR}_{1,2} = \frac{\text{Views of Video 1}}{\text{Views of Video 2}}$), represents the relative popularity of videos linked to Post 1 and Post 2. Values close to 1 indicate matched popularity. 
Here, paired post 1 and 2 are randomly assigned to avoid directional bias, making VVR symmetric. 
The Exact Phase is a special case within the Similar Phase, where VVR = 1.

Figure \ref{fig:score_views_interval} shows that as VVR deviates from 1, engagement increasingly favors posts linked to more popular videos, supported by a moderate Spearman correlation (0.299, $p < 0.001$). This highlights the utility of controlling video popularity effects without costly direct video analysis. We limit $\text{VVR}_{1,2}$ to a range of 0.5 to 2 (equivalent to undirected $\text{VVR}_{max}<2$). This criterion address the Exact Phase's limitations by expanding the pool of controlled data with stable timing and video similarity.

\paragraph{\InversePhase{} Phrase.} The final phrase investigates cases where the more popular post links to a less popular video, which reverses typical popularity expectations to emphasize the influence of titles. 
Specifically, we examine pairs where Score 1 > Score 2, but $\text{VVR}_{1,2} \leq 1$. These cases suggests that title content, rather than video popularity, is more likely to be the primary driver of engagement.

\subsection{Title Feature Selection and Analysis}

Building on the controlled framework, we refine the dataset using general confound control criteria, followed by text-specific filters.

For timing, we set a narrow time window of 0.5 hours for video posts. For video popularity, Similar cases use $0.5 \leq \text{VVR} \leq 2$. Exact cases default to $\text{VVR} = 1$, and Inverse cases default to a directed $\text{VVR}_{1,2} \leq 1$, with Post 1 being the more popular post.

To ensure meaningful title comparisons, we apply two thresholds based on Levenshtein Distance (LD), which measures textual similarity:
\begin{itemize}
    \item \textbf{Pairwise Similarity Threshold}: Pairs of titles with LD > 70, slightly above the dataset’s average LD of 63, are removed to preserve variation and prevent redundancy.  
    \item  \textbf{Video-Title Replication Threshold}: Titles closely replicate the original video title (LD > 95) are excluded. These minimal rewrites, comprising 21\% of all posts, often pair large score differences with negligible textual variation, obscuring the effects of title modifications.
\end{itemize}

Additionally, we exclude titles shorter than five characters or consisting of a single word. Consistent with prior work \cite{CatsAndCaptions,weissburg2022judging}, we require the more popular post to have at least double the score and a minimum score difference of 20 compared to its counterpart. This ensures meaningful contrasts in engagement levels.

Finally, the filtered post pairs from the Exact, Similar, and Inverse phases are consolidated into a unified Mixed dataset, with duplicates removed. Table \ref{tab:data_size} summarizes the final dataset sizes, which form the basis for the controlled analysis in the next subsection. To facilitate statistical tests, post pairs are ordered by engagement, with Post 1 being the more popular (Score 1 > Score 2), replacing earlier timing- or random-based assignments.


\begin{table}\setlength{\tabcolsep}{10pt}
\centering
\caption{Video post pairs across phases, refined with confounding controls and title-specific filters, for use in controlled analysis.}
\label{tab:data_size}
\begin{tabular}{ccccc} 
\toprule
\textbf{Time Window} & \textbf{Exact Match}  &\textbf{Similar Match} & \textbf{Inverse Match} & \textbf{Mix Dataset}  \\ 
\midrule
 0.5 hour & 1,883  & 36,501  & 67,359  & 86,035 \\
\bottomrule
\end{tabular}
\end{table}

\subsection{Title Features and Metrics}

We analyze title features across five categories: structural, lexical, stylistic, readability, and sentiment.  To ensure methodological rigor, these features are further divided into \textbf{continuous metrics}, which capture measurable properties, and \textbf{binary metrics}, which indicate the presence or absence of specific elements. 

\paragraph{Structural Metrics (Continuous).} These metrics capture basic composition and structure, including character count, word count, average word length, and average sentence length.


\paragraph{Lexical Diversity (Continuous).} Type-Token Ratio (\textbf{TTR}) measures the ratio of unique words to total words but decreases with longer texts due to word repetition.  To address this limitation, we prioritize Corrected TTR (\textbf{CTTR}) \cite{carroll1964language} and the Measure of Textual Lexical Diversity (\textbf{MTLD}) \cite{mccarthy2005assessment}, which adjust for text length and provide more reliable measurements for short titles. 


\paragraph{Stylistic Elements (Binary).}
Stylistic features influence title readability and emotional impact. 
Metrics include punctuation usage (exclamation marks, question marks, quotation marks), visual elements (emojis and numbers), and specific word types (pronouns, interrogatives, tentative words, certainty words, and affiliation words).   Additional stylistic patterns include fully uppercase words (at least three characters) for emphasis or shouting, and repeated characters (e.g., ``sooooo good,'' ``greeeat'', with at least three characters) to convey intensity.


\paragraph{Readability and Complexity (Continuous).} 
Readability metrics, such as Automated Readability Index (\textbf{ARI})~\cite{smith1967automated}, Coleman-Liau Index (\textbf{CLI})~\cite{antunes2019analyzing}, Flesch Reading Ease (\textbf{FR-Ease})~\cite{farr1951simplification}, Flesch-Kincaid Grade (\textbf{FK-Grade})~\cite{farr1951simplification}, and Gunning Fog Index (\textbf{G-F})~\cite{gunning1969fog},  are typically used to assess vocabulary difficulty and sentence structure in longer texts.
To ensure consistency, we reverse the FR-Ease scale so that higher scores consistently indicate greater complexity. 
For short titles, these metrics are complemented by structural and lexical diversity measures to mitigate biases and provide a more balanced analysis.

\paragraph{Sentiment and Emotional Metrics (Continuous and Binary).} 
We utilize both traditional lexicon-based methods and advanced deep learning models to analyze sentiment and emotion in titles. 


Traditional methods, such as \textbf{TextBlob} \cite{loria2018textblob}, \textbf{SentiWordNet} \cite{baccianella2010sentiwordnet}, and the \textbf{NRC Emotion Lexicon} \cite{mohammad2013crowdsourcing},  often struggle with short, informal social media text.  To enhance their applicability, we adapt them into binary metrics, assigning 1 if any sentiment or emotion (e.g., polarity, joy) is detected, regardless of type or intensity.

\textbf{VADER} is a lexicon-based method tailored for social media \cite{hutto2014vader}. Its ability to handle slang, emojis, and informal phrasing makes it more effective for analyzing social media titles.  
We use its continuous metrics, including sentiment proportions (positive, neutral, negative) and a composite score (compound) ranging from -1 (negative) to +1 (positive). 

To address the limitations of traditional methods, we use deep learning models trained on social network data.  Specifically, (i) \textbf{BERT-Sentiment Analysis}\footnote{https://huggingface.co/finiteautomata/bertweet-base-sentiment-analysis} predicts positive, negative, or neutral sentiment labels, and (ii) \textbf{BERT-Emotion Classification}\footnote{https://huggingface.co/bhadresh-savani/distilbert-base-uncased-emotion} identifies six emotions. We group them into positive (love, joy, surprise) and negative (fear, sadness, anger) categories.
Although BERT-Emotion does not explicitly handle neutral cases, we address this by comparing scores for high-level groups, where neutral texts yield balanced outputs near 0.5.  These advanced methods outperform traditional tools in handling the nuances of short, informal text.

\subsection{Statistical Methods and Setup}
To evaluate title features, we perform pairwise comparisons of metric differences between more popular posts (Post 1) and less popular posts (Post 2) in each pair. To address the challenges of analyzing social network data, including skewed distributions and confounding factors, we employed multi-dimensional statistical methods to ensure robust and reliable conclusions.  The choice of statistical tests was tailored to the type of metric being analyzed.

\paragraph{Continuous Metrics.} 
We employed both parametric and non-parametric tests.
The paired \textbf{\textit{t}-test}, a parametric test, compares the means of two paired samples under the assumption of normality. The Wilcoxon Signed-Rank Test (\textbf{W-test}), a non-parametric alternative, ranks absolute differences and focuses on directional trends, which is more resilient to outliers and skewed distributions. By employing both tests, we leveraged the \textit{t}-test’s strength for normally distributed data and the W-test’s robustness for non-normal datasets.

To assess normality, we used D’Agostino’s K-squared test, which revealed significant deviations from normality ($p < 0.001$) for most metrics. This is expected given the inherently skewed nature of social network data, patterns we have already analyzed in detail in Section \ref{sec:dataset}. While  achieving perfect normality is impractical in such datasets, the Central Limit Theorem supports the validity of the  \textit{t}-test for larger datasets. To ensure robustness, we interpret \textit{t}-test findings alongside W-test results and \textbf{rank biserial correlation} ($r_{rb}$) to quantify effect sizes. 
Typically, $r_{rb} \geq 0.4$ indicates a relatively strong practical difference, often supporting actionable insights.

To control for false positives across 22 tests, we applied a stricter Bonferroni correction, adjusting the significance threshold to $\alpha = 0.001/22$, which is even more stringent than $p < 0.001$.  Results meeting this threshold were considered robustly significant, while those near the threshold were interpreted cautiously in the context of their practical effect sizes.

\paragraph{Binary Metrics.}  For binary metrics, which capture the presence or absence of specific features, we applied the McNemar Significance Test (\textbf{M-test}). This test evaluates whether the proportions of binary features differ significantly between Group 1 and Group 2. Similar to our approach for continuous metric, we applied a Bonferroni correction ($\alpha = 0.001/16$) to account for multiple comparisons.

Unlike the \textit{t}-test and W-test, the M-test does not indicate the direction of differences. Thus, we also report Mean 1 and Mean 2, representing the proportions of titles in Group 1 and Group 2 that contain a given feature. Additionally, the Risk Difference (Mean 1 - Mean 2) quantifies how much more (or less) frequently a feature appears in Group 1 compared to Group 2, providing directional insight.

By integrating \textit{t}-tests, W-tests, $r_{rb}$, and M-tests, our analysis captures the complexities of social network data and ensures robust conclusions. Unlike prior studies relying solely on parametric tests, our multi-dimensional approach  provides a more reliable framework for analyzing title features and their impact on post popularity.

\subsection{Results and Analysis}
The results on the Mix Dataset, summarized in continuous metrics (Table \ref{tab:continous_result}) and binary metrics (Table \ref{tab:binary_result}), reveal distinct patterns across features. These findings underscore the nuanced role of title characteristics in shaping post popularity.  Titles that are informative, emotionally engaging, and strategically crafted drive higher engagement for video posts. Below, we present our findings.

\paragraph{Structural and Readability Metrics: Informative Titles Drive Engagement.}
In Table \ref{tab:continous_result},  Group 1 titles consistently exhibit greater character counts, word counts, and sentence lengths, with all metrics showing strong statistical significance (↑↑↑ \checkmark). These longer titles likely enhance informativeness, offering more context and detail to attract attention.

Readability metrics reinforce this observation.  Although originally designed for longer texts, they indicate that Group 1 titles are more complex. However, this complexity reflects informativeness rather than difficulty, as these titles remain accessible while being more descriptive and engaging. This trend challenges the conventional notion that ``simpler is better'' for titles, which may hold for general, longer text posts but does not necessarily apply to video posts. The observed differences suggest that audiences value titles that provide meaningful and specific information, aligning with the nature of video content where a compelling title can influence clicks and engagement.

\begin{table}[tbp]
\setlength{\tabcolsep}{10pt}
\centering

\caption{Statistical test results for continous metrics comparing Group 1 (more popular posts) and Group 2 (less popular posts) in the Mix Dataset. Paired \textit{t}-tests and Wilcoxon signed-rank tests (W-tests) are used, with significance levels indicated as: ↑↑↑ ($p<0.001$), ↑↑ ($p<0.01$), ↑ ($p<0.05$), and -- ($p>0.05$). Arrows (↑ or ↓) indicate whether Group 1 are larger or lower than Group 2. A checkmark (\checkmark) denotes results passing Bonferroni correction. Effect sizes are reported as rank biserial correlation ($r_{rb}$). 
\textbf{Conclusion} specifies whether Group 1 metrics are significantly larger or smaller than Group 2's,  or are \textbf{inconclusive} due to conflicting tests or weak effect sizes ($r_{rb}$).}
\label{tab:continous_result}

\begin{tabular}{crllrc} 
\toprule
\textbf{Group} & \textbf{Metrics}  & \multicolumn{1}{l}{\textbf{ \textit{t}-test }}  & \textbf{W-test}  & \multicolumn{1}{c}{\textbf{$r_{rb}$}} & \textbf{Conclusion}  \\ 
\midrule
\multirow{4}{*}{\rotatebox{0} {\textbf{Structural}}}  & \textbf{Chars}  & ↑↑↑ $\checkmark$  & ↑↑↑ $\checkmark$  & 0.40  & \multirow{4}{*}{Group 1 Larger}  \\
  & \textbf{Words}  & ↑↑↑ $\checkmark$  & ↑↑↑ $\checkmark$  & 0.37  & \\
  & \textbf{AvgWordL}  & ↑↑↑ $\checkmark$  & ↑↑↑ $\checkmark$  & 0.46  &   \\
  & \textbf{AvgSentL}  & ↑↑↑ $\checkmark$  & ↑↑↑ $\checkmark$  & 0.37  &  \\ 
\midrule
\multirow{3}{*}{\rotatebox{0}{\textbf{Lexical}}}  & \textbf{TTR}  & ↓↓↓ $\checkmark$  & ↓↓↓ $\checkmark$  & 0.05  & Inconclusive \\
  & \textbf{CTTR}  & ↑↑↑ $\checkmark$  & ↑↑↑ $\checkmark$  & 0.36  & \multirow{2}{*}{Group 1 Larger}   \\
  & \textbf{MTLD}  & ↑↑↑ $\checkmark$  & ↑↑↑ $\checkmark$  & 0.36  &  \\
\midrule
\multirow{5}{*}{\rotatebox{0}{\textbf{Readablity}}}  & \textbf{ARI}  & ↑↑↑ $\checkmark$  & ↑↑↑ $\checkmark$  & 0.46  & \multirow{5}{*}{Group 1 Larger}   \\
  & \textbf{CLI}  & ↑↑↑ $\checkmark$  & ↑↑↑ $\checkmark$  & 0.46  &  \\
  & \textbf{FK-Grade}  & ↑↑↑ $\checkmark$  & ↑↑↑ $\checkmark$  & 0.43  &   \\
  & \textbf{FR-Ease}  & ↑↑↑ $\checkmark$  & ↑↑↑ $\checkmark$  & 0.46  &   \\
  & \textbf{G-F}  & ↑↑↑ $\checkmark$  & ↑↑↑ $\checkmark$  & 0.42  &  \\ 
\midrule
\multirow{4}{*}{\rotatebox{0}{\begin{tabular}[c]{@{}c@{}}\textbf{Sentiment}\\\textbf{(VADER)}\end{tabular}}}  & \textbf{Positive}  & \multicolumn{1}{l}{\phantom{x}--} & \multicolumn{1}{l}{\phantom{x}--} & \multicolumn{1}{l}{\phantom{x}--}  & Inconclusive  \\
  & \textbf{Negative}  & ↑↑↑ $\checkmark$  & ↓↓↓ $\checkmark$  & 0.08  & Inconclusive \\
  & \textbf{Neutral}  & ↓↓↓ $\checkmark$  & ↓↓↓ $\checkmark$  & 0.24  & Group 1 Smaller  \\
  & \textbf{Compound}  & ↓↓↓ $\checkmark$  & ↓↓↓ $\checkmark$  & 0.25  & Group 1 Smaller  \\ 
\midrule
\multirow{4}{*}{\rotatebox{0}{\begin{tabular}[c]{@{}c@{}}\textbf{Sentiment}\\\textbf{(BERT)}\end{tabular}}}  & \textbf{Positive}  & ↑↑↑ $\checkmark$  & ↑↑↑ $\checkmark$  & 0.48  & Group 1 Larger  \\
  & \textbf{Neutral}  & ↓↓↓ $\checkmark$  & ↓↓↓ $\checkmark$  & 0.47  & Group 1 Smaller  \\
  & \textbf{Negative}  & ↑↑↑ $\checkmark$  & \phantom{x}-- & --\phantom{x} & Inconclusive  \\
  & \textbf{Biased}  & ↑↑↑ $\checkmark$  & ↑↑↑ $\checkmark$  & 0.47  & Group 1 Larger  \\ 
\midrule
\multirow{2}{*}{\rotatebox{0}{\begin{tabular}[c]{@{}c@{}}\textbf{Emotion}\\\textbf{(BERT)}\end{tabular}}}  & \textbf{Positive}  & ↓↓↓ $\checkmark$  & ↓↓↓ $\checkmark$  & 0.48  & Group 1 Smaller  \\
  & \textbf{Negative}  & ↑↑↑ $\checkmark$  & ↑↑↑ $\checkmark$  & 0.48  & Group 1 Larger  \\
\bottomrule
\end{tabular}
\end{table}

\begin{table}
\centering
\caption{Binary metrics comparison in the Mix phase: Results compare Group 1 (more popular posts) and Group 2 (less popular posts) using McNemar’s test.  Significance levels are denoted as  *** ($p<0.001$), ** ($p<0.01$), * ($p<0.05$), and -- ($p \geq 0.05$). A checkmark (\checkmark) denotes results passing Bonferroni correction. Risk Difference represents the absolute difference in proportions (Mean 1 - Mean 2).}
\label{tab:binary_result}
\begin{tabular}{clrrrr} 
\toprule
\textbf{Group}   & \textbf{Metrics} & \textbf{M-Test} & \textbf{Mean 1} & \textbf{Mean 2} & \textbf{Risk Diff} \\ 
\midrule
\multirow{12}{*}{\textbf{Stylistic}} 
  & \textbf{Exclamation Mark}   & ***$\checkmark$ & 5.8\%   & 5.1\%   & +0.7\% \\ 
  & \textbf{Question Mark}      & ***$\checkmark$ & 2.4\%   & 3.5\%   & -1.1\% \\ 
  & \textbf{Quotation Mark}	 & ***$\checkmark$	 & 4.0\%	 & 2.8\%	 & +1.1\%	 \\ 
  & \textbf{Numbers}       & ***$\checkmark$ & 36.9\%  & 34.3\%  & +2.6\% \\ 
  & \textbf{Emoji}         & ***$\checkmark$ & 5.9\%   & 5.1\%   & +0.8\% \\ 
  & \textbf{Uppercase}	 & ***$\checkmark$	 & 24.4\%	 & 25.7\%	 & -1.3\% \\ 
  & \textbf{Repeated Characters}	 & ***$\checkmark$	 & 2.1\%	 & 1.4\%	 & +0.6\%	 \\ 
  & \textbf{Pronouns}      & ***$\checkmark$ & 21.9\%  & 19.0\%  & +2.9\% \\ 
 & \textbf{Interrogatives}	 & --\phantom{xx}	 & 6.7\%	 & 6.5\%	 & +0.2\%	 \\ 
 &	\textbf{Tentative}	 & *\phantom{xx}	 & 1.0\%	 & 0.9\%	 & +0.1\%	 \\ 
 &	\textbf{Certainty}	 & ***$\checkmark$	 & 1.3\%	 & 1.1\%	 & +0.3\%	 \\ 
 &	\textbf{Affiliation}	 & ***$\checkmark$	 & 3.6\%	 & 3.2\%	 & +0.4\%	 \\ 
 
\midrule
\multirow{4}{*}{\textbf{Sentiment}} 
  & \textbf{TextBlob Subjectivity}  & ***$\checkmark$ & 49.5\%  & 45.0\%  & +4.5\% \\ 
  & \textbf{TextBlob Polarity}  & ***$\checkmark$ & 45.6\%  & 41.1\%  & +4.6\% \\ 
  & \textbf{SentiWordNet Polarity}  & ***$\checkmark$ & 60.6\%  & 56.9\%  & +3.7\% \\ 
  & \textbf{NRC Emotions} & ***$\checkmark$ & 18.5\%  & 16.5\%  & +2.0\% \\ 
\bottomrule
\end{tabular}
\end{table}

\paragraph{Lexical Diversity:  Adjusted Metrics Highlight Richer Vocabulary.}
Adjusted metrics like CTTR and MTLD, which account for text length, are significantly higher for Group 1, indicating richer vocabulary. In contrast, unadjusted metrics like TTR are slightly lower with negligible effect size 0.05, likely due to the longer length of Group 1 titles. This highlights the value of using length-adjusted metrics for short texts, as they reliably demonstrate that Group 1 titles are both longer and lexically richer.

\paragraph{Stylistic Features: Relatability and Professionalism Matter.}
Stylistic features, despite their infrequency, provide meaningful insights (Table \ref{tab:binary_result}). 
Pronouns (+2.9\%) stand out as effective in fostering a conversational tone and sense of community by directly addressing the audience. 
Numbers (+2.6\%) often appear as release dates, rankings like ``Top 10,'' or game scores, adding clarity and specificity that attract users seeking concise and actionable information.
Quotation marks (+1.1\%) highlight intriguing phrases or direct quotes, adding credibility and curiosity.

Conversely, some stylistic choices negatively impact engagement. 
Question marks (-1.1\%) suggest that open-ended questions may lack the specificity valued in video post titles. Similarly, uppercase words are significantly less common in Group 1 titles, even after accounting for length differences. This indicates that excessive emphasis or ``shouting'' through uppercase words may be perceived as unprofessional or spam-like, reducing their appeal.

Other stylistic elements, such as exclamation marks, emojis, repeated characters, certainty words, and affiliation words, show statistical significance but minimal risk differences. Features like interrogatives and tentative words show no significant differences, reinforcing the importance of strategically balancing relatability and professionalism in title crafting.

\paragraph{Sentiment and Emotional Metrics: Stronger Sentiment Attracts Engagement.}
Binary metrics adapted from traditional lexicon-based tools consistently reveal higher proportions of sentiment-related features in Group 1 titles (Table \ref{tab:binary_result}), reflecting stronger emotional content. While these tools cannot distinctly classify positive versus negative sentiment, they effectively capture the presence of emotional language.

Continuous sentiment metrics in Table \ref{tab:continous_result} provide more nuanced insights.
VADER, a lexicon-based model tailored for social media, shows no significant difference in positive scores and a negligible conflicting result for negative scores (effect size 0.08). However, it identifies Group 1 titles as less neutral, reflecting stronger overall negative sentiment in compound scores.
Deep learning models offer more  differentiation.  BERT sentiment analysis shows increases in both positive and negative sentiment for Group 1, with decreases in neutral sentiment. This suggests that both positive and negative sentiment drive higher engagement. 
BERT emotion classification further refines this observation, showing that negative emotions (e.g., sadness, anger) increase, while positive emotions (e.g., joy) decrease. This does not contradict the sentiment results but rather highlights the dual effectiveness of emotional appeals, where both positive and negative emotions can engage audiences depending on the context.

\begin{figure}[tbp]
  \centering
  \includegraphics[width=\linewidth]{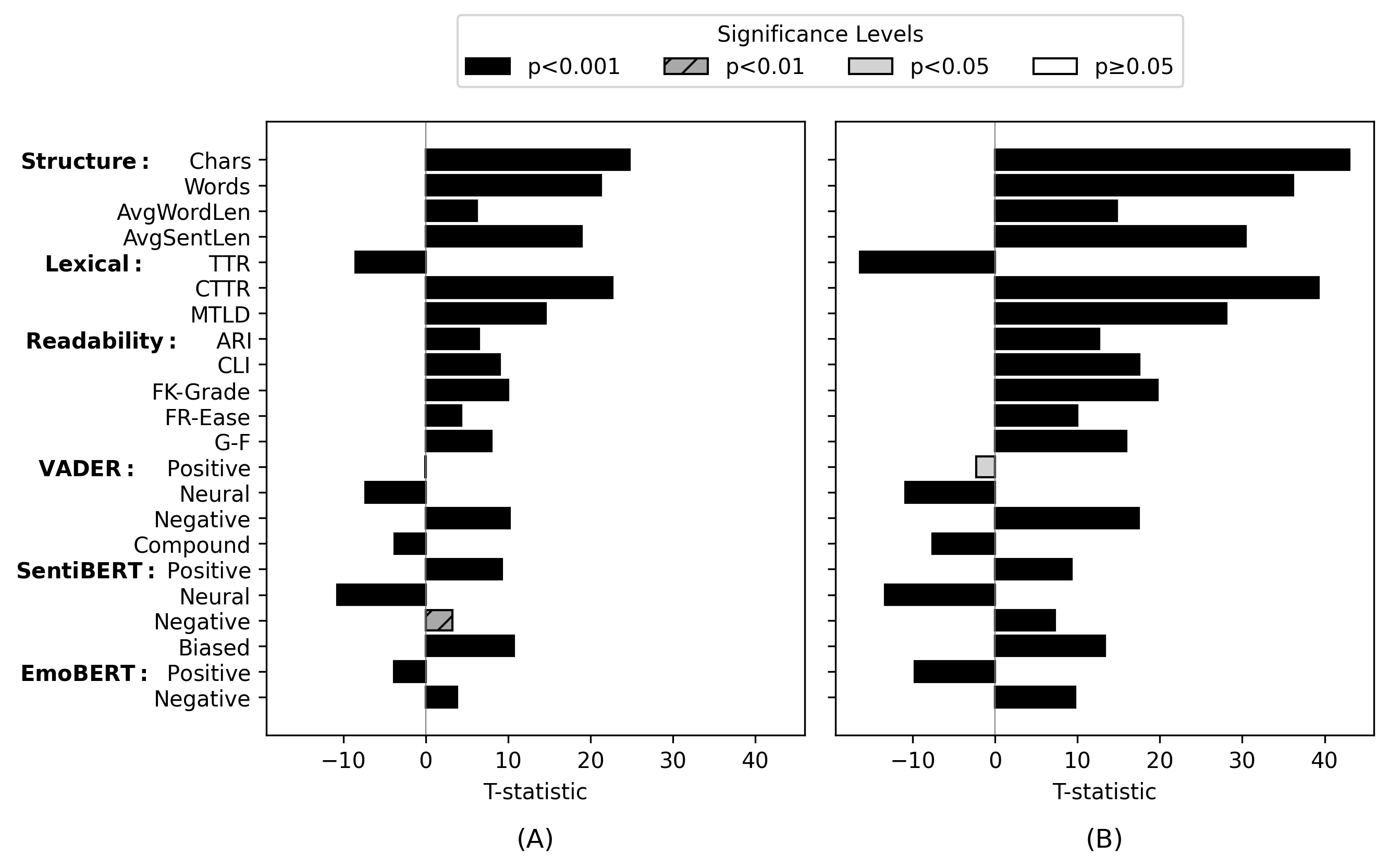}
  \caption{T-statistics from paired  \textit{t}-tests across Similar (A) and Inverse (B) phrases for continuous metrics. Bars represent the magnitude and direction of the standardized mean differences, with colors denoting significance levels. The amplified trends in Inverse cases highlight the increased influence of text features as video-related factors diminish.}
  \label{fig:t-test-similar-inverse}
  \Description{This figure displays two bar charts comparing t-statistics for continuous metrics across Similar and Inverse cases. Bars represent the magnitude and direction of standardized mean differences, with significance levels denoted by color. Inverse cases show intensified trends.}
\end{figure}

\paragraph{Amplified Text Effects in Inverse Match Phrase.}
Breaking down the results of continuous metrics on the Mix dataset into Similar and Inverse phases (Figure \ref{fig:t-test-similar-inverse}), we observe amplified trends in the Inverse phase. Paired \textit{t}-test T-statistics, which account for variability and sample size, highlight stronger effects in structural, readability, and lexical diversity metrics like CTTR and MTLD. These findings indicate that Group 1 titles in the Inverse phase emphasize informativeness and complexity more prominently. Sentiment and emotional metrics also show intensified patterns.
These amplified effects are not artifacts of sample size, as the larger sample size in the Inverse phase (67K vs. 36K in Similar) reduces variability and increases the reliability of the observed differences. The higher T-statistics suggest that textual features become more critical in driving engagement when video-related biases are minimized, highlighting the effectiveness of our experimental setup in isolating nuanced text-driven effects on post popularity.
\section{Controlled Prediction Experiments}\label{sec:Controlled Prediction}

\subsection{Prediction Task Setup}

While Section \ref{sec:Controlled Analysis} reveals how title characteristics influence post popularity, it does not fully capture the complexity of social media engagement. Prior studies show that even human predictions of post popularity barely surpass  random guessing \cite{CatsAndCaptions,weissburg2022judging}, highlighting the challenge of predicting popularity as many factors are subtle and context-dependent. 
Our study focuses on whether advanced language models can uncover patterns in text-based cues that elude human judgment, particularly within cross-platform, multimodal contexts.

To evaluate this, we design a pairwise ranking task \citep{ReinforcePopPredict,CatsAndCaptions,PairwisePopPredict1,weissburg2022judging}, predicting which of two posts $x_1$ or $x_2$ from a controlled post set $\mathcal{P}$ is more popular based solely on their titles.  This setup reduces the influence of confounding factors, focusing solely on the relative drivers of engagement.

\subsection{Baseline and Predictive Models}

Our experiments evaluate heuristic baselines, supervised deep learning models, and large language models (LLMs).

\paragraph{Heuristic baselines}
 These methods validate dataset controls and assess task complexity.  \textbf{Random Guessing} serves as a foundational baseline to confirm data balance. \textbf{Time-based ranking} assumes earlier posts receive higher engagement due to longer exposure, which validates our controls for timing effects. \textbf{Video views ranking} presumes that posts with more popular videos attract more engagement, testing our control over video influence.

\paragraph{Supervised Learning Models.}
We use \textbf{BERT-Base-Uncased} \cite{BERT} for contextualized text features, and a simpler Bidirectional \textbf{LSTM} \cite{hochreiter1997long} with FastText embeddings \cite{bojanowski2016enriching} as a baseline comparison. Both models are optimized using a max-margin loss, which encourages a higher score for the more popular post in each pair:

\begin{equation}
  \mathrm{Loss}(x_1, x_2)) = \mathrm{max}(0, x_2-x_1)
\end{equation}

Here, $x_i$ denotes the model's score for each post $i$. 

\paragraph{LLMs} 
We also explore \textbf{GPT-4o} \cite{openai2023gpt4} as an alternative to human judgment.  Human evaluations of post popularity often hover near random guessing and are impractical for large-scale annotation \cite{Tan14-Topic-and-author-controlled-Twitter,CatsAndCaptions,weissburg2022judging}. While not replacing human-level nuance, LLMs provide faster, scalable  alternative with consistent annotation performance \cite{gilardi2023chatgpt,jin2024better,hu2024leveraging}.  In our setup, GPT-4o receives title pairs and subreddit names, generating predictions and brief rationales (e.g., sentiment strength, relevance) to align its decisions with our statistical analysis. 

\subsection{Evaluation Dataset}
We derived training and testing splits from the controlled pairing datasets in Section \ref{sec:Controlled Analysis}. Specifically,  datasets were created with undirected $\text{VVR}_{max} < 2$ for Similar Match pairs and directed $\text{VVR}_{1,2} \leq 1$ for Inverse Match pairs, with a strict 0.1-hour time window. The Exact Match Phase was excluded  due to overlap with other phases. To ensure robust evaluation, we applied multiple data split strategies:

\begin{itemize}
  \item \textbf{Date-Based Split} (Default): Posts before 2022 for training and 2022 posts for testing. This split offers a realistic test setup, minimizing temporal overlap.
  \item \textbf{Post-ID Split}:  Pairs are assigned to the test set if either post ID belongs to the designated test group, preventing overlap in specific posts.
  \item \textbf{Video-ID Split}:  Pairs are assigned to the test set if either video ID falls in the test group, isolating video influence.
\end{itemize}

Both Post-ID and Video-ID splits used a 5\% test size to match the Date split, with multiple random seeds for statistical robustness. Sample sizes for one seed are presented in Table  \ref{tab:experiment_data_size}.

\begin{table}[htbp]\setlength{\tabcolsep}{8pt}
\caption{Data distribution across split strategies and phrases for train/test evaluation in controlled prediction experiments (example sizes shown for one random seed)}
\label{tab:experiment_data_size}
\centering
\begin{tabular}{ccccc} 
\toprule
\textbf{Split By}  & \textbf{Dataset Split} & \textbf{Similar Match} & \textbf{Inverse Match} & \textbf{Mix Dataset} \\ 
\midrule
\multirow{2}{*}{\textbf{Date}}  & \textbf{Train} & 9129 & 15393 & 20017 \\
  & \textbf{Test}  & 995  & 1506  & 2052  \\ 
\midrule
\multirow{2}{*}{\textbf{Post-ID}}  & \textbf{Train} & 9146 & 15208 & 19902 \\
  & \textbf{Test}  & 978  & 1691  & 2167  \\ 
\midrule
\multirow{2}{*}{\textbf{Video-ID}} & \textbf{Train} & 9122 & 15274 & 19925 \\
  & \textbf{Test}  & 1002 & 1625  & 2144  \\
\bottomrule
\end{tabular}
\end{table}

\begin{table}\setlength{\tabcolsep}{8pt}
\centering
\caption{Average accuracy results on the Mix test set across different data splits.}
\label{tab:Mix result}
\begin{tabular}{ccccccc} 
\toprule
\textbf{Phrase}  & \textbf{Random} & \begin{tabular}[c]{@{}c@{}}\textbf{Time}\\\textbf{Rank}\end{tabular} & \begin{tabular}[c]{@{}c@{}}\textbf{Video }\\\textbf{Views}\end{tabular} & \textbf{LSTM} & \textbf{BERT} & \textbf{GPT-4o}  \\ 
\midrule
\textbf{Date}  & 50.9  & 56.8  & 27.9  & 59.3  & 76.5  & 58.3  \\
\textbf{Post-ID}  & 50.1  & 52.8  & 23.3  & 57.5  & 72.5  & 59.5  \\
\textbf{Video-ID}  & 50.0  & 50.2  & 25.6  & 55.9  & 72.1  & 59.9  \\ 
\textbf{Average}  & 50.3  & 53.3  & 25.6  & 57.6  & 73.7  & 59.2  \\
\bottomrule
\end{tabular}
\end{table}

\begin{table}\setlength{\tabcolsep}{8pt}
\caption{Average accuracy results for different phrases within the Date-based split. This table expands upon the Date split results presented in Table \ref{tab:Mix result}.}
\label{tab:detailed phrase result}
\centering
\begin{tabular}{ccccccc} 
\toprule
\textbf{Phrase}  & \textbf{Random} & \begin{tabular}[c]{@{}c@{}}\textbf{Time}\\\textbf{Rank}\end{tabular} & \begin{tabular}[c]{@{}c@{}}\textbf{Video }\\\textbf{Views}\end{tabular} & \textbf{LSTM} & \textbf{BERT} & \textbf{GPT-4o}  \\ 
\midrule 
\textbf{Similar} & 51.3  & 58.0  & 57.5  & 52.8  & 73.4  & 59.9  \\
\textbf{Inverse} & 50.8  & 56.3  & 1.7  & 60.3  & 76.4  & 56.2  \\
\textbf{Mix}  & 50.9  & 56.8  & 27.9  & 59.3  & 76.5  & 58.3  \\
\bottomrule
\end{tabular}

\end{table}

\subsection{Prediction Results}

Table \ref{tab:Mix result} summarizes average accuracy results across splits (Date, Post-ID, Video-ID) on the Mix test set, averaged over multiple random seeds. 
Table \ref{tab:detailed phrase result} provides phase-specific results (Similar, Inverse, and Mix) for the Date split.  We omit phrase-specific results for Post-ID and Video-ID splits as they follow similar trends.

\paragraph{Split Robustness and Baseline Performance.} Model performance remains consistent across splits, which reinforces dataset robustness and the reliability of our evaluation framework. As expected, random guessing yields around 50\% accuracy across all splits and phrases. Time-based ranking offers slight improvements but remains limited, validating our timing controls. Video views ranking performs poorly on Inverse pairs, confirming their neutrality toward video influence. These baselines highlight the inadequacy of single factors like timing or video popularity in predicting engagement.

\paragraph{GPT-4o Performance.} GPT-4o marginally outperforms random guessing, reflecting the inherent complexity of predicting post popularity from short titles.  The task requires capturing nuanced contextual cues that are challenging to infer, even for models with advanced reasoning capabilities. While prior studies indicate that human judgment faces similar difficulties, GPT-4o highlights both its potential and limitations. It generates plausible rationales, such as familiarity (noted in 49\% of cases), sentiment strength, relevance to current events, and nostalgia/exclusivity (appeals to memory or unique access). However, it fails to consistently capture contextual nuances essential for accurate predictions in short-title, cross-platform settings.

\paragraph{Supervised Learning Models.} BERT achieves approximately 74\% accuracy across splits and phrases, underscoring the importance of contextualized text representation. This strong performance also validates our statistical findings, highlighting the role of nuanced textual features in post popularity. In contrast, LSTM struggles due to limited contextual understanding,  despite overfitting mitigation measures such as freezing embeddings, dropout, and regularization. 
Both models perform better on Inverse Match phrases, consistent with our findings in Section \ref{sec:Controlled Analysis} that these settings amplify text patterns.

In summary, these results highlight the complexity of predicting social media engagement in a cross-platform, multimodal context. Individual factors are insufficient predictors. However, our controlled multi-phrase setup effectively manages confounding factors, allowing advanced models to capture nuanced text patterns that often elude human judgment, revealing deeper insights into engagement dynamics.

\subsection{Further Insights and Robustness Tests}

We tested the Pairwise Similarity Threshold to determine the optimal level for pairing titles. The threshold is set at LD = 70, excluding overly similar pairs (LD > 70) to preserve meaningful variation. Our ablation study revealed a strong negative correlation (-0.824) between the threshold and test accuracy, indicating that overly similar paired titles (high LD) confuse the model and hinder performance. LD = 70 strikes a balance, allowing the model to capture meaningful distinctions while maintaining data size.

We also test the necessity of the Video-Title Replication Threshold, which excludes titles that closely replicate the original video title (LD > 95). Without this filter, model accuracy drops to 65\%, highlighting its importance.  This threshold need not be overly restrictive, as its primary purpose is to remove minimal rewrites. Such pairs often feature large score differences with negligible textual variation, obscuring the effects of title modifications.  While this trade-off the dataset size by 21\%, it significantly enhances data quality—a priority given the inherent complexity and randomness of social data.

We experimented with incorporating more text features—such as subreddit names, original YouTube video titles, and post dates—by concatenating them with the post titles.  However, this did not consistently improve accuracy (less than 1\%) across all seeds or splits, suggesting that title rewriting alone is the primary factor influencing post performance when confounding factors are controlled.

Additionally, we explored a range of BERT-based architectures, including distilled, optimized, and robust variants \cite{sanh2019distilbert,liu2019roberta,hu2022conflibert}, and tested both cased and uncased settings,  but observed no significant performance differences.
The large performance gap between LSTM and BERT-based models highlights the importance of contextual embeddings, suggesting that BERT-level contextualized models are sufficiently robust for this task. Larger models, however, risk overfitting due to inherent noise in engagement prediction. Prior work \cite{weissburg2022judging} also demonstrates that simplified attention-based architectures achieve comparable results to BERT, underscoring that modeling limits lie in the task's inherent difficulty rather than model complexity.

A notable challenge in predicting relative popularity is the inherent engagement randomness. Even highly similar posts still show different scores due to this randomness. Grouping similar posts and retaining the maximum, mean, or median scores could be a potential solution, but challenges remain, especially in determining how to group posts with varying modification styles. Nevertheless, the increasing dataset size helps improve the reliability of our predictions, demonstrating they are consistently more accurate than chance.

\subsection{Explaining Prediction Dynamics}

Building on prediction results, we analyze attention patterns and case studies to reveal how textual features drive engagement.

\subsubsection{Attention Analysis Across Subreddits.}
We analyzed attention distribution across subreddits to explore how text is valued differently across communities. To enhance interpretability, fragmented subwords from BERT’s tokenization were reconstructed into full words.  Since attention scores are context-specific, we focused on the \textbf{most frequently highlighted words}, counting those with the highest attention weight in each sentence. Although frequent words may be slightly overrepresented, this approach effectively reveals linguistic patterns and community preferences. 
Despite no subreddit-specific fine-tuning, BERT’s attention aligns with unique community characteristics.

Table \ref{tab:attention_word_list} shows the top 10 most frequently highlighted words across subreddits, uncovering thematic preferences.  For example, \texttt{r/Music}, \texttt{r/kpop}, and \texttt{r/90sHipHop} emphasize musical genres, iconic artists, and cultural elements, reflecting  music-oriented communities’ cultural and artist-driven focus. In contrast, sports subreddits like \texttt{r/nba} prioritize player names and basketball-related terms  (e.g., ``lebron,'' ``dunks"), while persona-driven subreddits like \texttt{r/SquaredCircle} highlight key figures like ``cena'' and narrative terms like ``promo''.

Subreddits centered on emotional or sensory engagement, such as \texttt{r/asmr} and \texttt{r/youtubehaiku}, prioritize terms like ``roleplay,'' ``tingly,'' and ``poetry.''  reflecting their community's emphasis on creative and immersive experiences. Meanwhile, gaming subredditsl ike \texttt{r/Games} reveal a preference for technical language, including references to trailers and gaming events. 

By highlighting words' contextual importance, attention mechanisms provide complementary insights to frequency-based methods like TF-IDF, revealing nuanced interpretations of community norms.

\begin{table}
\centering
 \caption{Top 10 most frequently highlighted words by attention across subreddits}
  \label{tab:attention_word_list}
\begin{tabular}{ll} 
\toprule
\textbf{Subreddit} & \multicolumn{1}{c}{\textbf{Highlighted Words (by Frequency)}}  \\ 
\midrule
r/Music  & indie, grunge, pop, halen, ska, faith, wave, rap, depeche, cure  \\
r/kpop  & dreamcatcher, loona, gfriend, teaser, weeekly, mamamoo, wjsn, dle, txt, chungha  \\
r/nba  & lebron, nba, westbrook, embiid, harden, kawhi, raptors, dunks, bryant, steph  \\
r/youtubehaiku  & haiku, poetry, meme, reddit, anakin, bongo, conga, darth, minaj, oc  \\
r/Hololive  & hololive, pekora, gura, suisei, holostars, miko, sora, aki, fubuki, haachama  \\
r/SquaredCircle  & spoilers, wwe, aew, nxt, ago, promo, cena, njpw, raw, strowman  \\
r/90sHipHop  & 2pac, wu, goodie, nas, outkast, efx, nature, cube, jeru, mobb  \\
r/asmr  & asmr, roleplay, intentional, binaural, frivolousfox, inaudible, ozley, portugues, unintentional, 60fps  \\
r/Games  & e3, ps4, xbox, trailer, alaloth, gamescom, foundry, naheulbeuk, tga, valhalla  \\
\bottomrule
\end{tabular}
\end{table}

\begin{figure}
    \centering
    \includegraphics[width=1\linewidth]{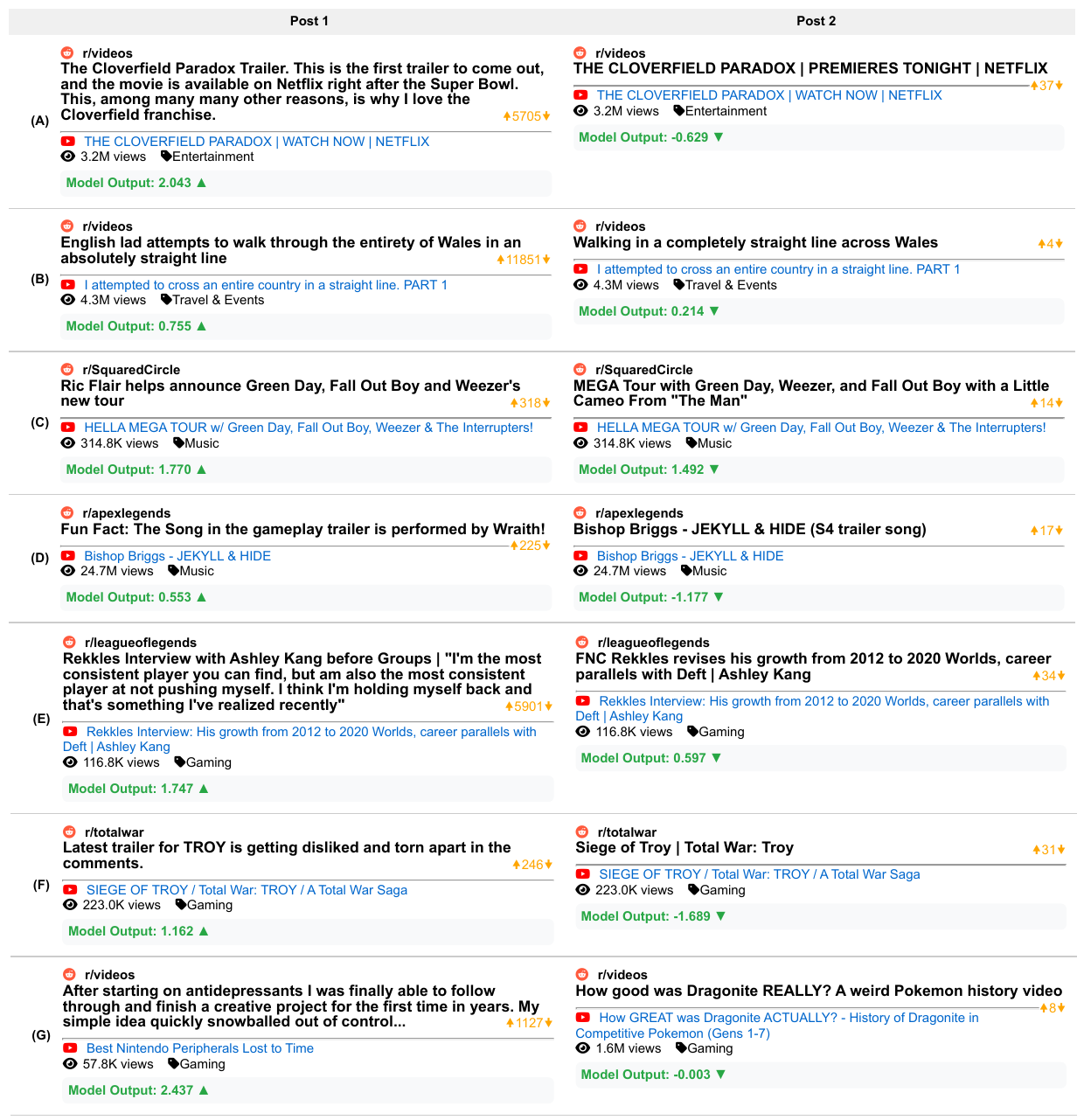}
    \caption{Case studies of title effectiveness in post pairs: Post pairs (A–H) from the controlled dataset compare more popular (Post 1) and less popular (Post 2) posts, shown with real post scores (orange) and model predictions (green). Subreddit names, titles, and video details (titles, views, and categories) are included. Arrows indicate which post the model predicts as more engaging.}
    \label{fig:real_example}
    \Description{This figure displays case studies of title effectiveness for pairs of posts from the controlled dataset. Each pair shows the title, score, and model predictions, with arrows indicating which post the model predicts as more engaging. Examples highlight the role of emotional appeal, curiosity, and subreddit relevance in engagement.}
\end{figure}

\subsubsection{Case Studies of Title Effectiveness}
To complement the quantitative findings, we selected seven representative post pairs from our controlled prediction dataset, as shown in Figure \ref{fig:real_example}. Each pair consists of a more popular Post 1 and a less popular Post 2, with post scores (orange) and model predictions (green) illustrating real and predicted engagement differences. We primarily selected exact match pairs (same video for both posts) to ensure engagement differences are attributed to title variations rather than video content. Additionally, Pair H represents an inverse match case, where Post 1 achieved higher engagement despite linking to a less popular video, highlighting the influence of textual features over video views.

The observed patterns align with findings from Section \ref{sec:Controlled Analysis}. For instance, the effectiveness of structural and readability features (e.g., longer, more informative titles), lexical richness, and emotional or stylistic elements emerges as a consistent theme. Below, we analyze these pairs in detail.

Pair A exemplifies the power of emotional, personalized titles. Post 1 aligns with Reddit’s conversational norms, offering personal insight and emotional engagement. In contrast, Post 2 provides no additional value beyond the video title itself.  This example reinforces the importance of informative and emotionally resonant titles.

Pair B demonstrates the value of curiosity and clarity. Post 1 intrigues readers with a title that emphasizes novelty. In contrast, Post 2 is vague and fails to clearly communicate the video’s content. This aligns with our findings that more descriptive and engaging titles drive better outcomes.

Pair C showcases the impact of subreddit-specific relevance.  
Both posts discuss Green Day’s tour, but Post 1 incorporates ``Ric Flair,'' a culturally significant figure within \texttt{r/SquaredCircle}, enhancing resonance with the subreddit audience.    
The model assigns high scores to both posts but gives a higher score to Post 1, reflecting its ability to understand subreddit-specific influencers. Interestingly, the real post scores show a clear disparity, illustrating the ``rich-get-richer'' effect: Post 1's catchy title likely captured early visibility, overshadowing Post 2 despite their close publication times.

In Pair D, Post 1’s emotional and narrative-driven title outperformed Post 2’s direct replication of the video title. While subreddit rules like ``Artist - Song Title'' may limit title variations, emotional or informative additions can still lead to better engagement. This reinforces the potential of thoughtful rewrites even within rule-based constraints.

Gaming-related pairs (E-G) reveal diverse strategies.  In Pair E, an emotionally charged title resonates strongly with the subreddit’s community, outperforming a more generic, news-like tone. In Pair F, Post 1’s creative framing of controversy captures attention more effectively than Post 2’s straightforward description. Lastly, Pair G underscores the influence of textual features over external factors. Despite Post 1 linking to a less popular video, its engaging and heartfelt narrative struck a chord with the audience. These examples collectively emphasize how personal and emotional elements can often outweigh more conventional metrics, such as video popularity, in driving engagement.

These case studies reinforce the broader trends identified in our controlled analysis. Informative, lexically rich, and emotionally resonant titles consistently outperform their counterparts across diverse subreddits and topics. While certain community norms (e.g., \texttt{r/Music}’s formatting rules) may constrain rewrites, these constraints reflect real-world conditions rather than methodological flaws. Our pairwise approach mitigates such limitations by focusing on meaningful textual differences, offering actionable insights into the dynamics of title effectiveness on Reddit.

\section{Conclusion}
Our study advances the understanding of cross-platform content engagement by systematically investigating how title rewrites influence user interactions in Reddit posts sharing YouTube videos.  Through a multi-phase, controlled experiment, we neutralized confounding factors such as video popularity, timing, and community norms to isolate the effects of textual variations. The findings demonstrate that title rewrites significantly impact engagement, revealing patterns like longer titles, sentiment shifts, and community-specific keywords that drive user interactions.

While effective in isolating textual effects, our controlled experimental setup may not fully capture the complexity of real-world interactions, including ecosystem-wide dynamics and evolving user behavior. Additionally, the study’s focus on textual features overlooks the complementary role of multimodal elements such as video thumbnails or descriptions. Finally, restricting the dataset to Reddit-YouTube interactions may limits the generalizability of our findings to other platform pairs with distinct user behaviors and norms.

Despite these limitations, this study offers actionable strategies for social media practitioners and a replicable framework for researchers. By tailoring titles to audience preferences, incorporating emotional resonance, and optimizing informativeness, practitioners can enhance engagement with minimal effort. The framework’s design for isolating confounding factors makes it adaptable to other cross-platform contexts. Testing this approach on diverse platform pairs, such as Twitter-Instagram or TikTok-YouTube, would validate its applicability and uncover additional insights into engagement dynamics.

Future research should incorporate multimodal features such as video content, thumbnails, and metadata to provide a more comprehensive understanding of engagement. Advanced attention mechanisms from state-of-the-art models can further explore how users prioritize content features, improving both analysis and predictive accuracy. Sophisticated causal inference techniques tailored to multimodal and cross-platform settings could refine the study of textual and visual interactions. Expanding datasets to include a wider variety of platforms with distinct user demographics and sharing norms would also enhance the robustness and generalizability of future findings.

\begin{acks}
This research/material is based upon work supported in part by NSF grants CNS-2154118, ITE-2137724, ITE-2230692, CNS2239879, Defense Advanced Research Projects Agency (DARPA) under Agreement No. HR00112290102 (subcontract No. PO70745), CDC, and funding from Microsoft. Any opinions, findings, and conclusions or recommendations expressed in this material are those of the author(s) and do not necessarily reflect the position or policy of DARPA, DoD, SRI International, CDC, NSF, and no official endorsement should be inferred. We thank members of the SocWeB Lab and CLAWS Lab for their helpful feedback.
\end{acks}


\bibliography{reference}
\bibliographystyle{ACM-Reference-Format}

\end{document}